# Uncertainty Estimation in Multi-Agent Distributed Learning for AI-Enabled Edge Devices

Gleb Radchenko[1][a] and Victoria Andrea Fill[2][b]
*[1] Silicon Austria Labs, Sandgasse 34, 8010 Graz, Austria*
gleb.radchenko@silicon-austria.com
*[2] FH Joanneum, Alte Poststraße 149, 8020 Graz, Austria*
victoria.fill@edu.fh-joanneum.at

Abstract: Initially considered as low-power units with limited autonomous processing, Edge IoT devices have seen a paradigm shift with the introduction of FPGAs and AI accelerators. This advancement has vastly amplified their computational capabilities, emphasizing the practicality of edge AI. Such progress introduces new challenges of optimizing AI tasks for the limitations of energy and network resources typical in Edge computing environments. Our study explores methods that enable distributed data processing through AI-enabled edge devices, enhancing collaborative learning capabilities. A key focus of our research is the challenge of determining confidence levels in learning outcomes, considering the spatial and temporal variability of data sets encountered by independent agents. To address this issue, we investigate the application of Bayesian neural networks, proposing a novel approach to manage uncertainty in distributed learning environments.

## 1 INTRODUCTION

Traditionally, Internet of Things (IoT) Edge devices have been perceived primarily as low-power components with limited capabilities for autonomous operations (Samie et al., 2016). However, in recent years, the focus of IoT research has shifted towards optimizing knowledge exchange and implementing AI on edge devices. These advancements are largely due to the innovation of FPGAs and AI accelerators, which have exponentially increased the computational capabilities of Edge devices (Liang et al., 2023; Parmar et al., 2023; Wang et al., 2022).

This evolution raises critical questions that system developers should address:

- **Knowledge Exchange:** How can we implement seamless knowledge sharing between edge devices to refine machine learning algorithms while maintaining data privacy?
- **Resource Management:** What strategies can effectively manage the computational power of these increasingly autonomous, high-performance devices?
- **Spatiotemporal Locality:** How can we address the localized nature of data to ensure real-time or near-real-time task execution?

The challenges presented by limited resources on edge devices and the spatiotemporal locality of data are particularly significant. These issues require new approaches to manage computational capabilities and efficiently perform tasks in real-time or near-real-time modes.

The goal of this research is to investigate the algorithms and methods for deploying distributed machine learning within the framework of autonomous, network-capable, sensor-equipped, AI-enabled edge devices. Specifically, we focus on determining confidence levels in learning outcomes, considering the spatial and temporal variability of data sets encountered by independent agents. To address this issue, we investigate the potential of the Distributed Neural Network Optimization (DiNNO) algorithm (Yu et al., 2022), aiming to extend it for organizing distributed data processing and uncertainty estimation using Bayesian neural networks.

---

[a]  https://orcid.org/0000-0002-7145-5630
[b]  https://orcid.org/0009-0006-6289-0098

Within the scope of this paper, we explore the interaction of AI-enabled edge devices using the case of robotic platforms engaged in the task of collaborative mapping. To achieve this, we need to address the following tasks:

- Implement a simulation of robots navigating a 3D environment using the Webots platform (Michel, 2024), augmented with advanced LiDAR sensors for detailed environmental mapping.
- Decouple the DiNNO algorithm implementation into independent processes, enabling asynchronous network communication for distributed learning.
- Integrate distributed uncertainty estimation into the resulting models by applying Bayesian neural networks.

The rest of the paper is structured as follows. Section 2 is devoted to analyzing the state-of-the-art research, providing an overview of distributed machine learning methods, and implementing Bayesian Neural Networks for uncertainty estimation. In Section 3, we present a collaborative mapping case. Section 4 introduces a distributed implementation of the DiNNO framework. Section 5 focuses on distributed uncertainty estimation, exploring techniques for integrating BNN into DiNNO. Implementation details and the evaluation of our approaches are detailed in Section 6, followed by Section 7, which offers our conclusions and discussion of future work directions.

## 2 RELATED WORK

### 2.1 Distributed Machine Learning Methods

Distributed machine learning (ML) algorithms, distinguished by their communication mechanisms, primarily support the exchange of model parameters, model outputs, or hidden activations. These exchanges can be enabled through peer-to-peer or client-server architectures (Park et al., 2021). The primary approaches utilized within these algorithms may be categorized as follows.

#### 2.1.1 Federated Learning

**Federated Learning** (FL) orchestrates the periodic transmission of local training parameters (e.g., weights and gradients of a neural network) from workers to a central parameter server. This server then performs model averaging and disseminates the updated global model to the workers. Such a strategy not only may preserve data privacy by avoiding the need for raw data exchange but also may enhance communication efficiency through adjustable transmission intervals (Lim et al., 2020). The authors of (Abreha et al., 2022) identify FL as a solution to challenges in edge computing environments, such as unwanted bandwidth loss, data privacy issues, and legalization concerns. They highlight that FL allows for co-training models across distributed clients, such as mobile phones, automobiles, and hospitals, via a centralized server while maintaining data localization.

The authors of (Nguyen et al., 2022) propose an extension of the FL model, called FedFog, designed to enable FL over a wireless fog-cloud system. The authors address key challenges such as non-identically distributed data and user heterogeneity. The FedFog algorithm performs local aggregation of gradient parameters at fog servers and a global training update in the cloud.

#### 2.2.2 ADMM-derived Methods

**Alternating Direction Method of Multipliers (ADMM)-derived Methods** (Boyd, 2010) (such as DiNNO (Yu et al., 2022), GADMM, and CADMM) aim to implementation of distributed learning in the absence of a central coordinating entity by enabling communication directly between the neighboring worker nodes in a peer-to-peer (P2P) manner. One critical issue of FL and such P2P learning methods is that the communication overhead is proportional to the number of model parameters, limiting their efficacy in supporting deep neural networks (Elgabli et al., 2020).

#### 2.1.3 Federated Distillation

**Federated Distillation** utilizes the exchange of model outputs, which are significantly lower in dimensionality compared to the full model sizes for distributed learning. In this approach, each worker performs local iterations based on its individual loss function. This process is enhanced with a regularization component that measures the discrepancy between the worker's predicted output for a specific training sample and the aggregated global output for the same class. A widely known application of knowledge distillation is model compression, through which the knowledge of a large pretrained model may be transferred to a smaller one (Ahn et al., 2019).

### 2.1.4 Split Learning

**Split Learning** (SL) partitions multi-layer neural networks into segments, thus making it possible to train large-sized deep NN that exceed the memory capacities of a single edge device. This approach divides the NN into *lower NN segments* on the workers' devices, each containing raw data and a shared *upper NN segment* hosted on a parameter server. The *NN cut layer* is a boundary between the lower and upper NN segments. Workers compute the activations at the NN cut layer and send these activations to the parameter server. The parameter server uses these activations as inputs for the upper NN segment to continue the forward pass, compute the loss, and initiate the backward pass. Gradients calculated at the cut layer are then transmitted back to the workers, allowing them to update the weights of the lower NN segments. However, the efficacy of SL in terms of communication is subject to ongoing discussion (Koda et al., 2020).

### 2.1.4 Multi-agent Reinforcement Learning

**Multi-agent Reinforcement Learning** extends beyond traditional regression or classification objectives to address scenarios where environmental dynamics influence worker decisions. In such contexts, workers must learn these dynamics and adapt their strategies based on the knowledge acquired through interactions with the environment and among themselves (Buşoniu et al., 2008).

## 2.2 Uncertainty Estimation and Bayesian Neural Networks

### 2.2.1 Bayesian Neural Networks

In a conventional neural network architecture, a linear neuron is characterized by a weight ($w$), a bias ($b$), and an activation function ($f_{act}$). Given an input $x$, a single linear neuron performs the following operation:

$$y = f_{act}(w \cdot x + b) \quad (1)$$

where $y$ is the output of the neuron.

Bayesian Neural Networks (BNNs) employ a Bayesian approach to train stochastic neural networks (Jospin et al., 2022). Instead of deterministic weights and biases, they utilize probability distributions, denoted $P(w)$ for weights and $P(b)$ for biases. Typically, these distributions are approximated as Gaussian, with mean and standard deviation derived from the training data. Hence, a Bayesian neuron outputs a range of possible values, not just one. So, the operation of a Bayesian Linear neuron can be described as:

$$P(y|x) = f_{act}\left(\sum P(w) \times x + P(b)\right) \quad (2)$$

In a BNN, the Gaussian distributions for weights and biases may be defined by the mean $\mu$ and the standard deviation $\sigma$. For weights, the distribution $P(w)$ is modeled as a Gaussian distribution with a mean $w_\mu$ and a standard deviation $w_\sigma$, where:

$$w_\sigma = \log(1 + e^{w_\rho}) \quad (3)$$

The parameter $w_\rho$ ensures that the standard deviation is always positive. Similarly, the distribution $P(b)$ for biases is represented as a Gaussian distribution with a mean $b_\mu$ and a standard deviation $b_\sigma$, where:

$$b_\sigma = \log(1 + e^{b_\rho}) \quad (4)$$

During the forward pass of a Bayesian neuron, these distributions are sampled to obtain a weight and bias for each neuron. The sampled weights and biases are then used to compute the neuron's output. The parameters $w_\mu$, $w_\rho$, $b_\mu$, $b_\rho$ are learned during NN training to optimize the network's performance.

Unlike conventional NNs, which employ a singular forward pass, BNNs might conduct multiple forward passes. The mean and standard deviation of these outputs are then computed. Depending on the problem type the neural network addresses, these mean and standard deviation values can indicate the model's uncertainty for each point in the input data space.

### 2.2.2 Kullback-Leibler Divergence

Kullback-Leibler Divergence (KL Divergence) (Claici et al., 2020; Kullback & Leibler, 1951) is employed to account for the difference between the Gaussian distributions that represent the parameters of the BNN. KL Divergence serves as a measure to quantify the dissimilarity between two probability distributions and can be generally computed as:

$$D_{KL}(g \parallel h) = \int g(x) \log \frac{g(x)}{h(x)} dx \quad (5)$$

where $g(x)$ and $h(x)$ are two probability density functions defined over the same support. The concept of "expected excess surprise" captures the core idea behind KL Divergence, reflecting the expected degree of "surprise" encountered when another "model" distribution approximates an actual distribution. As outlined by (Belov & Armstrong, 2011), if $N_0(\mu_0, \sigma_0)$ and $N_1(\mu_1, \sigma_1)$ are two normal

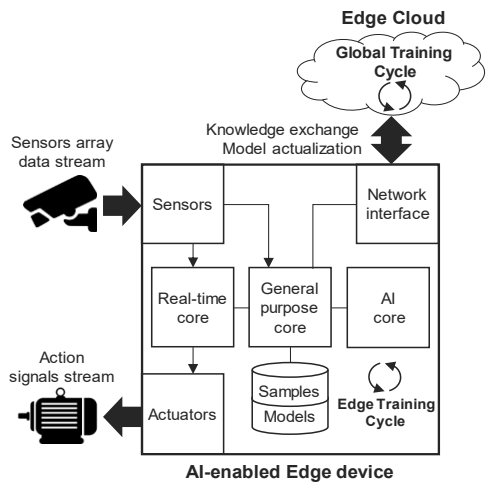

Figure 1: Components of AI-Enabled edge device.

probability density functions, equation (5) may be reduced to:

$$D_{KL}(N_0 \parallel N_1) = \frac{1}{2}\left(\log\frac{\sigma_1^2}{\sigma_0^2} + \frac{\sigma_0^2 + (\mu_0 - \mu_1)^2}{\sigma_1^2} - 1\right) \quad (6)$$

Within the BNNs, applying KL Divergence helps quantify the deviation of the neural network's parameter distribution from a specified prior distribution. The overall loss in a BNN model is generally expressed as:

$$total_{loss} = base_{loss} + kl_{weight} \times kl_{loss} \quad (7)$$

where $base_{loss}$ refers to the conventional loss function, such as Binary Cross-Entropy or Mean Squared Error; $kl_{weight}$ is a hyperparameter that enables adjusting of the level of uncertainty in the model's outcomes; and $kl_{loss}$ is the sum of the KL Divergence between the distribution of BNN modules $N_0(\mu_0, \sigma_0)$ and a predefined normal distribution $N_1(\mu_1, \sigma_1)$.

## 3 COLLABORATIVE MAPPING CASE

As a case study for a distributed AI application operating within the Edge Cloud, we have chosen a collaborative environment mapping problem. This task involves deploying a network of independent, robotic edge devices (robots) at various starting points. Each device is tasked with building a coherent map of the environment, utilizing installed sensors, and exchanging knowledge about the environment with other devices.

These devices are designed to update a local ML model with newly acquired data samples and facilitate inter-device communication via a network interface (see Figure 1). Equipped with computational cores dedicated to specific responsibilities, the devices feature:

- **Real-time core** for immediate data processing and direct control of actuators.
- **General-purpose core** for overall device control.
- **AI core** to support an edge training cycle.

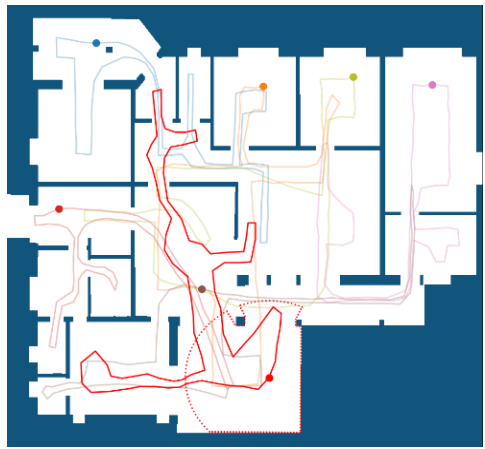

Figure 2: Visualization of the environment map, including starting points and exploration pathways for the robotic agents, as described by (Yu et al., 2022).

The Distributed Neural Network Optimization (DiNNO) (Yu et al., 2022) algorithm is employed as the principal method for addressing the distributed machine learning problem in our study. DiNNO enables the implementation of distributed learning within a network of independent agents. These agents are identical robotic platforms regarding computational capabilities and sensor equipment. Each robot possesses its own sensor data set and maintains an individual version of the neural network (NN). These robots refine their NN models throughout the learning phase using fresh sensory inputs and then exchange NN parameters. This iterative process ensures that, over time, all agents align on a harmonized NN representation. The CubiCasa5K data set (Kalervo et al., 2019) was used as a reference for the floor plans generation (see Figure 2).

## 4 DISTRIBUTED EDGE LEARNING APPROACH

To optimize the DiNNO algorithm for edge computing environments, it is essential to transition from a centralized learning framework that relies on sequential learning processes based on shared agents' memory. Consequently, a distributed implementation of the DiNNO framework was designed, wherein each agent functions as an independent process, with communication achieved through message exchanges. Each agent implements the local LiDAR data processing, local optimization of the NN parameters, and message-based exchange of the updated NN parameters with their peers.

We have introduced an epoch-based decentralized consensus algorithm to support the decentralized peer-to-peer exchange of NN parameters among agents (see Algorithm 1). The maximum amount of synchronization epochs (*MaxRound*), network socket (*Socket*), unique peer identifier (*Id*), and the initial state of the NN parameters (*State*) are given as inputs for the algorithm. *PeerComplete[]* and *PeerState[]* structures are utilized to track the completion status and states of peers, respectively.

The core of the algorithm lies in the exchange of two types of messages:
- The *RoundComplete* message indicates the completion of a round by a peer.
- The *State* message contains the peer's state for the current round.

The introduction of the *RoundComplete* message alongside the round finalization logic addresses the issues introduced by the latency in the message delivery. These challenges include out-of-order messages, delayed status updates, and desynchronization between rounds. A *RoundComplete* message is sent by a peer only after it has received all *State* messages from the other peers. It ensures that a peer only advances to the next round once all peers have completed the current round, indirectly handling message delays by waiting for all messages to be received before proceeding.

As for the out-of-order messages, the agent checks if the received *State* message is from the future round. If so, it triggers the FINISHROUND function to ensure the peer catches up to the correct round. This mechanism helps in managing out-of-order deliveries due to latency.

This version of the algorithm operates under the assumption that each message sent will eventually be received by its intended recipient. In this context, we do not account for scenarios involving agent malfunctions or halts, nor do we consider the permanent failure of network equipment that could lead to the irreversible loss of messages or the complete breakdown of communication between peers.

---

**Algorithm 1. Peers State Exchange**
**Require:** *MaxRound, Socket, Id, State*
**Initialize:** *Round, PeerComplete*[ ], *PeerState*[ ]
*Message* ← (*State*, 0)
SEND(*Socket*, *Message*, *Id*)
**while** *Round* < *MaxRound* **do**
    (*Message, PeerId*) ← RECEIVE(*Socket*)
    **if** *Message* is *RoundComplete* **then**
        *PeerComplete*[*PeerId*] ← TRUE
    **else**
        **if** *Round* < *Message.Round* **then**
            FINISHROUND
        **end if**
        *PeerState*[*PeerId*] ← *Message.State*
    **end if**
    **if** $\forall s \in PeerState, s \neq \emptyset$ **then**
        *State* ← NODEUPDATE(*State, PeerState*)
        $\forall s \in PeerState, s \leftarrow \emptyset$
        *PeerCompleted*[*Id*] ← TRUE
        *PeerState*[*Id*] ← *State*
        *Message* ← *RoundComplete*
        SEND (*Socket, Message, Id*)
    **end if**
    **if** $\forall p \in PeerComplete, p =$ TRUE **then**
        FINISHROUND
    **end if**
**end while**
**function** FINISHROUND
    $\forall p \in PeerComplete, p \leftarrow$ FALSE
    *Round* ← *Round* + 1
    *Message.State* ← *State*
    *Message.Round* ← *Round*
    SEND (*Socket, Message, Id*)
**end function**

Algorithm 1: Peers State Exchange.

## 5 DISTRIBUTED UNCERTAINTY ESTIMATION

To address uncertainty estimation in the distributed mapping problem, we incorporate a BNN by replacing the conventional linear layers in the neural network with Bayesian Linear Layers. The architecture of the BNN is detailed as follows (see Figure 3):

- *Input Layer (2):* x, y – an input coordinate representing the global position on the environment map.
- *SIRENLayer (256):* a layer with a sinusoidal activation function suitable for Neural Implicit Mapping.

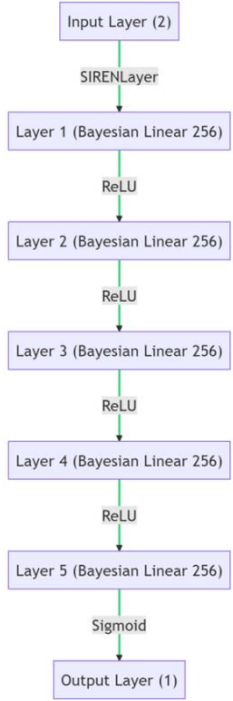

Figure 3: Proposed Bayesian Neural Network architecture.

- *4 x Bayesian Linear Layers (256):* four Bayesian Linear layers with 256 nodes each, activated by the ReLU function. These layers are probabilistic and support uncertainty estimation.
- *Output Layer (1):* a linear layer with one node activated by the Sigmoid function.

This modification introduces probabilistic inference to the model, allowing for estimating uncertainty in the network's predictions.

One of the main advantages of BNNs is their ability to provide a measure of uncertainty in their predictions. Unlike deterministic networks, which offer a single output value, BNNs can perform multiple forward passes to compute the mean and standard deviation of the outputs. This provides a measure of the model's uncertainty for each point in the mesh grid, which can be invaluable in applications where understanding the model's confidence level is crucial.

To ensure correct regularization of the BNN parameters during the distributed learning regularization phase, Algorithm 2 has been developed to consider the semantics of median ($\mu$) and standard deviation ($\rho$) parameters of BNN neurons. We utilize KL Divergence, as detailed in Equation (6), for the regularization of BNN $\rho$-parameters between the models of individual actors.

---

Algorithm 2. Optimization of BNN Parameters

**Require:** *Model, Optimizer$_\mu$, Optimizer$_\rho$, W$_\mu$, W$_\rho$, Iter, $\theta^\mu_{reg}$, $\theta^\rho_{reg}$, Duals$_\mu$, Duals$_\rho$*
**for** $i \leftarrow 1$ to *Iter* **do**
    Reset gradients of *Optimizer$_\mu$* and *Optimizer$_\rho$*
    *PredLoss* ← COMPUTELOSS(*Model*)
    $\theta^\mu, \theta^\rho$ ← EXTRACTPARAMETERS(*Model*)
    *Reg$_\mu$* ← L2REGULARIZATION($\theta^\mu, \theta^\mu_{reg}$)
    *Reg$_\rho$* ← D_KL($\theta^\rho, \theta^\rho_{reg}$)
    *Loss$_\mu$* ← *PredLoss* + $\langle\theta^\mu, Duals_\mu\rangle$ + $W_\mu \times Reg_\mu$
    *Loss$_\rho$* ← $\langle\theta^\rho, Duals_\rho\rangle$ + $W_\rho \times Reg_\rho$
    UPDATEPARAMETERS(*Optimizer$_\mu$, Loss$_\mu$*)
    UPDATEPARAMETERS (*Optimizer$_\rho$, Loss$_\rho$*)
**end for**

Algorithm 2: Optimization of BNN Parameters.

## 6 IMPLEMENTATION AND EVALUATION

Based on floor plans sourced from the CubiCasa5K dataset, we generated 3D interior models in STL format for robotic exploration. To simulate the behavior of autonomous agents, these 3D interior models were imported into the Webots simulation platform (Michel, 2024), where we deployed models of TurtleBot robots for navigation within these environments (see Figure 4). Webots support implementing of 3D models of robotic systems and environments, including simulating the most typical sensors, such as cameras and LiDAR. This methodology enabled us to use advanced LiDAR sensor models, incorporating realistic noise and measurement uncertainties into our experiments.

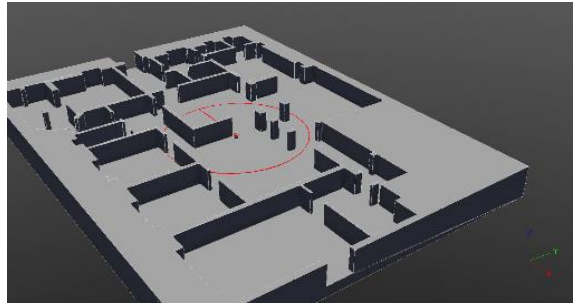

Figure 4: Visualization of the 3D model of the environment, generated from the floor plan, showcasing a LiDAR-equipped TurtleBot navigating the space in a Webots simulation.

In this study, it is assumed that all robots can access global positioning information. Movement paths for the agents were pre-determined, enabling the generation of simulation programs for their traversal through the interiors. As agents navigate, information gathering via LiDAR sensors is simulated as a data stream produced by the Webots process.

The experiment involves launching seven independent agents that gradually collect information from LiDAR sensors while exploring a virtual interior space. The source of training data can be either a pre-recorded file containing results from LiDAR data collections or a live data stream from LiDAR sensors obtained during the Webots simulation. This paper explores the results derived from analyzing data sets collected and processed collectively after the agents finished their traversal. This approach is necessitated by the substantial processing power and energy resources required for neural network training, which may not be available to autonomous edge devices operating in a mobile investigation mode.

Each agent runs as a separate Python process. Agent communication is handled through direct TCP connections among the processes within the same virtual local network. The ZeroMQ library is used for asynchronous data exchange. Containerization of agent processes is achieved using Singularity containers equipped with GPU access to ensure a stable simulation and data analysis environment. In the experiments outlined, we initiate all processes on GPU-enabled computing nodes managed by the SLURM workload manager. The methodology for this multi-process learning experiment is visually summarized in Figure 4.

## 6.1 Single-Agent Uncertainty Estimation

We conducted a series of experiments with a single isolated agent to evaluate the effectiveness of the BNN architecture proposed in Section 5 for estimating uncertainty in neural network outcomes and the impact of the $kl_{weight}$ parameter from Equation (7). The agent was trained exclusively on local data during the experiment without exchanging information with other agents. The visualization of the training results is presented in Figure 5.

To generate outputs from the Bayesian neural network, 50 queries were made for each pair of input coordinates $(x, y)$. Subsequently, a visualization was created to illustrate the mean values and standard deviations of the neural network responses.

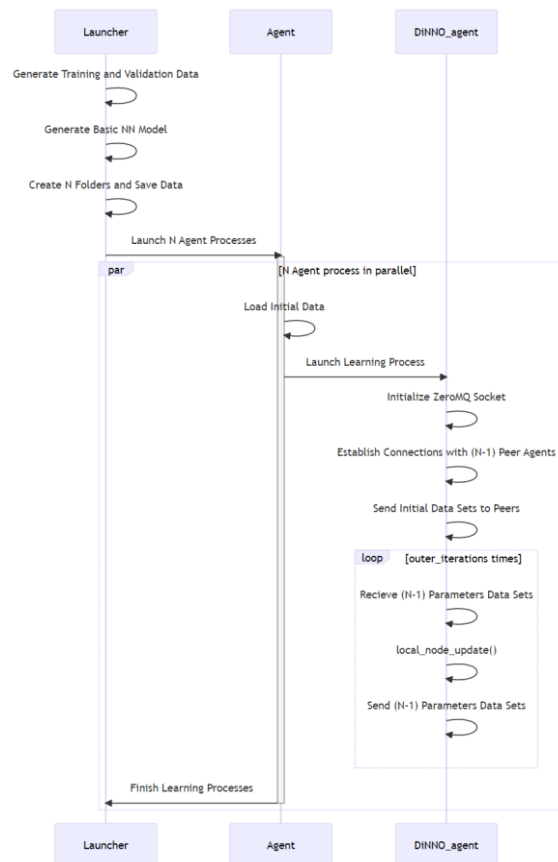

Figure 5: Sequence diagram of multi-process learning experiment.

It was observed that a low value of the $kl_{weight}$ parameter leads to a low variance in the neural network's results, which does not allow for distinguishing the "hallucinations" of the neural network from areas with sufficient data to form a general understanding of the environment. Conversely, a high $kl_{weight}$ parameter value results in excessive noise and a high degree of uncertainty in the neural network's results. Therefore, to ensure that the Bayesian neural network provides an effective assessment of uncertainty, fine-tuning of the $kl_{weight}$ parameter during the training process is required.

## 6.2 Multi-Agent Uncertainty Estimation

To assess the effectiveness of Bayesian neural networks in estimating uncertainty within distributed,

(a) Mean, $kl_{weight} = 10^{-4}$

(b) Mean, $kl_{weight} = 5 \times 10^{-3}$

(c) Mean, $kl_{weight} = 5 \times 10^{-1}$

(d) Standard deviation, $kl_{weight} = 10^{-4}$

(e) Standard deviation, $kl_{weight} = 5 \times 10^{-3}$

(f) Standard deviation, $kl_{weight} = 5 \times 10^{-1}$

Figure 6: Comparative visualization of single-agent uncertainty estimation.

decentralized learning environments, a series of experiments were conducted. These experiments aimed to evaluate the impact of different regularization approaches on the training quality of Bayesian neural networks. The validation loss was evaluated in the context of the following regularization strategies:

- Uniform L2 regularization of NN parameters without making distinctions between parameter types;
- Separate regularization of conventional and Bayesian neural network parameters, applying L2 regularization for both;
- Separate regularization of conventional and Bayesian neural network parameters, utilizing L2 regularization for conventional parameters and Kullback–Leibler divergence for Bayesian parameters (see Algorithm 2).

The results of the evaluation are presented in Figure 6. We observe that applying Kullback–Leibler divergence for parameter regularization (Algorithm 2) leads to a 12-30% decrease in the validation loss of the distributed BNN training compared to other regularization strategies. Additionally, this approach enhances the stability of the training process. The visualization of the outcome of decentralized BNN training according to Algorithm 2 is presented in Figure 8.

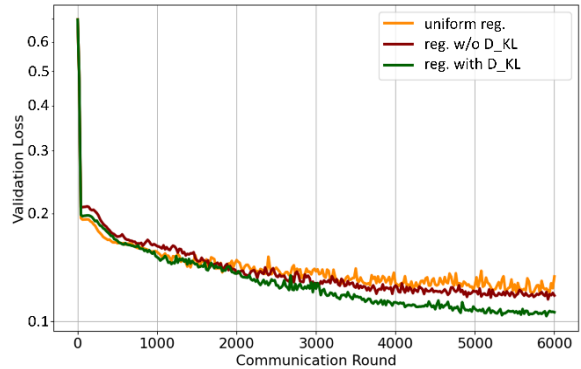

Figure 7: Comparison of validation loss during distributed BNN training 1) with uniform L2 regularization (*uniform reg.*); 2) separate L2 regularization (*reg. w/o D_KL*); 3) Kullback-Leibler divergence for regularization of BNN ρ-parameters (*reg. with D_KL*).

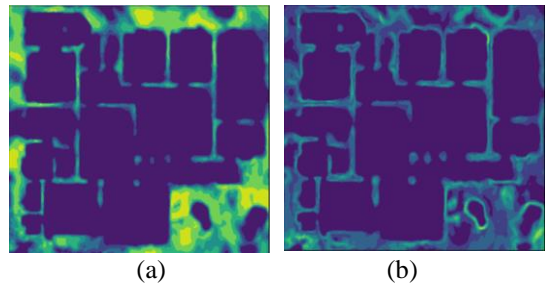

(a)          (b)

Figure 8: Visualization of the decentralized BNN training results according to Algorithm 2: a) mean; b) standard deviation.

# 7 CONCLUSIONS

Within the scope of this paper, we addressed a problem of uncertainty estimation within distributed machine learning based on AI-enabled edge devices. We set up a simulation of a collaborative mapping problem using the Webots platform; introduced an epoch-based decentralized consensus algorithm to support the decentralized peer-to-peer exchange of NN parameters among agents; and integrated distributed uncertainty estimation into our models by applying Bayesian neural networks.

Our experiments indicate that BNNs can effectively support uncertainty estimation in a distributed learning context. However, to ensure an effective assessment of uncertainty by the BNN, we highlighted the need for precise tuning of the learning hyperparameters during training. We also determined that applying Kullback–Leibler divergence for parameter regularization resulted in a 12-30% reduction in validation loss during the training of distributed BNNs compared to other regularization strategies.

For future work, we suggest exploring how distributed learning with BNNs can be tailored for embedded AI hardware. This would involve refining the NN architecture to suit the resource constraints of AI-enabled edge devices. We also plan to explore task management and offloading strategies within the multi-layered fog and hybrid edge-fog-cloud environments to improve computational efficiency and resource utilization.

# ACKNOWLEDGMENTS

The research reported in this paper has been partly funded by the European Union's Horizon 2020 research and innovation program within the framework of Chips Joint Undertaking (Grant No. 101112268). This work has been supported by Silicon Austria Labs (SAL), owned by the Republic of Austria, the Styrian Business Promotion Agency (SFG), the federal state of Carinthia, the Upper Austrian Research (UAR), and the Austrian Association for the Electric and Electronics Industry (FEEI)